\documentstyle[12pt,epsfig]{article}
\begin{document}
\baselineskip= 1.5pc
\begin{centerline}
{TWO STEP MECHANISM OF $\eta, \ \eta'\ ,\omega, \ \phi $ MESON}
\end{centerline}
\begin{centerline}
{PRODUCTION IN  $pD\to ^3HeX$ REACTION}
\end{centerline}
 \begin{center}
L.A. Kondratyuk \footnote {e-mail address:kondratyuk@vxitep.itep.msk.su} \\
{\footnotesize
\em Institute of Theoretical and Experimental Physics,\\ \footnotesize \em
B.Cheremushynskaya 25, Moscow 117259,Russia}

 Yu.N. Uzikov \footnote {e-mail address: uzikov@nusun.jinr.dubna.su\\
 {\it Former address}: Department
of Physics, Kazakh State University, Timiryazev Str.,47,
Almaty 480121 Republic of Kazakhstan\\
}  \\

\footnotesize \em Laboratory of Nuclear Problems,\\
\footnotesize \em Joint Institute for Nuclear Research, Dubna,
Moscow reg. 141980, Russia
\end{center}
\section*{Abstract}
The differential cross sections of $pD\to ^3He X$ reactions, where
$X=\eta, \ \eta'\ ,\omega, \ \phi $, are calculated on the basis
of a  two-step mechanism
involving the  subprocesses $pp\to d\pi ^+$ and $\pi^+n\to Xp $.
It is shown that this model describes well the form of energy
dependence of available experimental cross sections at the final c.m.s.
 momentum  $p^*$=0.4-1$ GeV/c$ for the $\eta $
and at $p^*=0-0.5$ GeV/c for the $\omega $ meson  as well as  the ratios
 $R(\eta'/\eta) $ and $R(\phi/\omega)$.  The absolute value of the cross
section
is underestimated by an  overall normalization factor of  about 3 for $\eta, \
\eta '$ and
by nearly one order of magnitude  for $\omega$ and $\phi$. The spin-spin
correlations are predicted for  the reaction ${\vec D}{\vec p}\to ^3He X$
in the forward-backward approximations for the elementary amplitudes.
\bibliographystyle{unsrt}
\vspace{1.5cm}

PACS: 25.10.+s, 25.40.Qa, 13.60.Le, 14.40.Aq

KEYWORDS: $pd$ interaction, $\eta,\ \eta', \ \omega, \ \phi $ meson production.

\newpage

1. Reactions $pD\to ^3He X$, where $X$ means a meson heavier than the pion,
are of great interest for several reasons. Firstly, high momentum transfer
  ($\sim 1$ GeV/c) to the nucleons takes place in these processes.
 Secondly,
unexpected strong energy dependence of $\eta $ meson production was observed
near the threshold \cite{berg88}. In this respect the possible  existence
 of quasi-bound
states in the  $\eta -^3He$ system is discussed in the literature
\cite{kuwilk}-
\cite{kugree}.
Thirdly,  production of the $\eta, \eta', \phi $ mesons, whose wave functions
contain valence strange quarks, raises a question concerning strangeness
 of the nucleon and the mechanism of Okubo-Zweig-Iizuka rule violation
 \cite{kuellis},\cite{kuwurz}.
As a result of this discussion
the experimental investigation of the reaction ${\vec D} {\vec p}\to ^3He\phi$
is proposed \cite{kukhac} in Dubna.
Finally, the preliminary experimental data on $\eta '$ ¨ $\phi $ meson
production
in the reactions $pD\to ^3He \eta'$ and  $pD\to ^3He \phi$ near the thresholds
 are available at present  \cite {kusaturn} at meson c.m.s. momenta
 $p^*\sim 20\ MeV/c$.  New experimental data on the $pD\to ^3He \omega $
 reaction  were obtained recently in Ref. \cite {kuwurz95}
 above the threshold. In this connection
 the mechanism of the reactions in question seems to be very important.

2. The first  attempt to describe the reaction
 $pD\to ^3He \eta $ on the basis of the three-body
 \cite{kulagel} mechanism displayed an important role of
 intermediate pion beam in this process.
As  was mentioned for the first time in Ref. \cite{kukilian}, at the
 threshold  of
the reaction $pD\to ^3He \eta $ the two-step mechanism including two
subprocesses
$pp\to d \pi ^+$ and $\pi^+n\to \eta p $ is favoured. The advantage of this
mechanism is that at the threshold of this reaction and zero  momenta of Fermi
motion in the deuteron and  $^3He$ nucleus the amplitudes of these subprocesses
are practically on the energy shells. It is easy to check, that this
 peculiarity (the so called velocity matching or kinematic miracle)
 takes place above the threshold too, if the c.m.s. angle $\theta_{c.m.}$
of the  $\eta $ meson production  in respect
to the proton beam is $\theta_{c.m.}\sim 90^o$. For the $\omega, \ \eta'$ and
$\phi $ mesons the velocity
matching takes  place above the corresponding thresholds only at
$\theta_{c.m.}\sim 50^o-90^o$ depending on the meson mass and energy of
the incident proton.

 The two-step model of the
  $pD\to ^3He \eta $ reaction was developed in Refs. \cite{kuklu},
 \cite{kuwilkf}. Progress in comparison with the microscopic model
\cite{kulagel}
 was the comprehension of the very important role of final state interaction
 in the $\eta -^3He $ system near the threshold.
  Recently  F\"aldt and Wilkin \cite{kuwilkfpl} found that the two-step
 model can describe the form of the threshold cross section of $pD\to ^3He X^0$
 reaction as a function of the mass of produced meson  $X^0=
 \eta, \omega, \eta',\phi $. New points of present work are following.
 (i) We extend the two-step model \cite{kuklu} for the production of $\eta,
\omega,
 \eta'$ and $\phi $ mesons above the thresholds ( at the final c.m.s momenta
 $p^*$ about several hundred MeV/c). Previously, only $\eta $ meson production
 was  investigated above the threshold by Laget and Lecolley in the microscopic
  model.
  (ii) We consider in more general form than in Ref. \cite {kuwilkfpl}
 the influence the spin effects. In particular, we predict the pD spin-spin
 correlation for the reaction ${\vec D}{\vec p}\to ^3He X$ at the energy
 region of the proposed  Dubna experiment \cite {kukhac}.

3. Proceeding from the 4-dimensional technique of
nonrelativistic graphs
one gets the following expression for the amplitude of the  $pD\to ^3He X $
reaction in the framework of the two-step model corresponding to the Feynman
graph in Fig.1
 Ref.\cite {kuklu}
\begin{equation}
 A(pD\to ^3He X)= C{\sqrt 3\over{ 2m}} A_1(pp\to d\pi^+)
 A_2(\pi^+n\to Xp) {\cal F}(P_0,E_0),
\label{1}
\end{equation}
where  $A_1$ and $A_2$ are the amplitudes of the $pp\to d\pi^+$
and $\pi^+n\to Xp$ processes respectively, $m$ means the nucleon mass,
  $C=3/2$ is the isotopic spin factor taking into account the sum over
isotopic spin projections in the intermediate state. This factor is  the
same for all isoscalar mesons  $X$ under discussion. The form factor
has the form
\begin{equation}
\label{2}
{\cal F} (P_0,E_0)=\int {d^3q_1\over {(2\pi )^3}} {d^3q_2\over {(2\pi )^3}}
{\Psi_D({\bf q}_1) \Psi_{\tau }^*({\bf q}_2)\over {E_0^2-({\bf P}_0+{\bf
q}_1+{\bf q}_2)^2
+i\epsilon}}.
\end{equation}
 Here $\Psi_D({\bf q}_1)$ is the deuteron wave function
and $ \Psi_{\tau }({\bf q}_2)$ is the $^3He$ wave function in momentum space
for the  $d+p$ channel; $E_0$ and ${\bf P}_0$ are the energy and momentum
of the intermediate $\pi $ meson at zero Fermi momenta in the nuclear vertices
 ${\bf q}_1={\bf q}_2=0$:
\begin{equation}
 E_0=E_X+{1\over 3}E_{\tau }-{1\over 2}E_D, \ \
{\bf P}_0=-{2\over 3}{\bf P}_{\tau } - {1\over 2} {\bf P}_D,
\label {3}
\end{equation}
 where $E_j$ is the energy of the j-th particle in c.m.s., ${\bf P}_D$ and
 ${\bf P}_{\tau }$ are the relative momenta in the initial and final states
respectively $|{\bf P}_{\tau }|\equiv p^*$.
In comparison with \cite{kuwilkf} we do not restrict ourselves to the linear
 approximation over ${\bf q}_1$ and ${\bf q}_2$ in the $\pi $ meson
propagator and take into account the dependence on Fermi momenta exactly.
It results in faster decreasing  $|{\cal F} (P_0,E_0)|$ with increasing
 mass of the meson produced than in Ref.\cite {kuwilkf}.

Amplitude
  (\ref{1}) is related to the differential cross section of the
 $pD\to ^3He X$ reaction by the following expression
\begin{equation}
\label {4}
{d\sigma\over {d\Omega }}={1\over{64\pi ^2}}{1\over s_{pd}}
{|{\bf P}_{\tau }|\over {|{\bf P}_D|}} {\overline {|A(pD\to ^3He X)|^2}}
={|{\bf P}_{\tau }|\over {|{\bf P}_D|}}|f(pD\to ^3HeX)|^2,
\end{equation}
 where $\sqrt{s_{pD}}$ is the  invariant p+D mass. The amplitudes
$A_1(pp\to d\pi^+)$ and $ A_2(\pi^+n\to X) $ are similarly related
 to the corresponding cross sections. When deriving  Eq.
(\ref {1} ) one factored the amplitudes of elementary subprocesses
 $A_1$ and $A_2$ outside the integral sign over
 ${\bf q_1}$ and ${\bf q_2}$ at the point ${\bf q}_1={\bf q}_2=0$ and then
 replaced them to the  amplitudes of the corresponding free processes.
Neglection of the off-energy-shell effects is expected to be correct at
 the velocity matching
 conditions. Taking into account the off-shell and Fermi motion effects
in the optimal approximation \cite{kugurvitz}  one obtains numerical results
very close to the
approximation  (\ref {1})  if the energy dependence of the
cross sections   of  elementary processes is smooth enough.

The cross section can be always present  in the following formally separable
form
\begin{equation}
\label{5}
{d\sigma\over {d\Omega }}=R_S K |{\cal F}(P_0,E_0)|^2 {d\sigma\over {d\Omega }}
(pp\to d\pi^+) {d\sigma\over {d\Omega }}(\pi^+n\to Xp)
\end{equation}
 where $K$ is the kinematic factor defined according to Eq. (21) in Ref.
\cite{kuklu} for the differential cross section developed in a spinless
approximation.
( Indeed the factor $K$ from Ref.\cite {kuklu} is multuplied  here by factor
 $(9/8)^2$ in order  to obtain the correct  normalization condition for the
 the vertex  function $d+p\to ^3He$).

  The additional factor $R_S$ in Eq.(\ref{5}), which is absent
in Ref.\cite{kuklu}, takes into account spins and generally depends on
 mechanism of the reaction.
 It is important to remark that  the approximation  (\ref {1})
 does not lead  generally to the condition $R_S=const$
because of complicated spin structure of the amplitudes
$A_1(pp\to d\pi^+)$ and $A_2(\pi^+n\to Xp)$.
 The analysis is simpler at the angles $\theta _{c.m.}=0^o$ and $ 180^o$.
In this case the production of pseudoscalar meson
$\pi^+ n\to Xp$ in the forward-backward direction is described by only one
 invariant amplitude. The processes $pp\to d\pi^+$ and
 $\pi^+n\to \omega(\phi )p$
 are determined by two forward-backward invariant amplitudes
 $a_i$ and $b_i$   according to the following expressions
 \cite {kugermw90}
\begin{equation}
\label{6}
{\hat A_1(pp\to d\pi^+}) = a_1 {\bf e n} +i b_1 {\bf \sigma } [{\bf e}\times
{\bf n}],
\end{equation}
\begin{equation}
\label{6a}
{\hat A_2(\pi^+n\to p\omega }) = a_2 {\bf e  \sigma} + b_2
({\bf \sigma  n})({\bf e  \sigma }),
\end{equation}
where  ${\bf n}$ is  the unity vector along the incident proton beam,
 ${\bf e}$ is the polarization vector of the spin 1 particle
$(d, \omega ,\phi, )$, ${\bf {\sigma }}$ denotes the Pauli matrix.
According to our numerical calculations, the contribution of the $D-$
component of the nuclei wave functions  to the  modulus squared of the form
factor $|{\cal F}(P_0,E_0)|^2$ is less than $\sim 10$ \% for
 the deuteron and  less $\sim 1$ \% for $^3He$.
Using the $S-$ wave approximation for
the nuclear wave functions and taking into account Eqs.(\ref{6},\ref{6a})
we have found
the following expressions for the  spin factor $R_S$ of the spin averaged
cross section in the two-step model
\begin{equation}
\label{7}
R_{0 }={1\over 3}\left ( {1\over 2}|a_1|^2+{2\over 3}|b_1|^2-{2\over 3}
Re(a_1b_1^*)\right )
\left [{1\over 2}|a_1|^2+|b_1|^2\right ]^{-1}
\end{equation}
-- for the pseudoscalar mesons and
$$R_{1 }={1\over 3}\left [{1\over 2}|a_1|^2(3|a_2 |^2+\gamma )+{2\over 3}
(|a_2 |^2+\gamma )Re(a_1b^*_1)+{2\over 3}|b_1|^2(5|a_2|^2+\gamma )\right ]$$
\begin{equation}
\label{7a}
\times \left [{1\over 2}(|a_1|^2+2|b_1|^2)(3|a_2|^2+\gamma)
\right ]^{-1}
\end{equation}
for the  vector mesons,
where $\gamma= |b_2|^2+2Re(a_2^*b_2)$.
As it follows from Ref.
\cite{kugermw90}, at the threshold of
$\eta $ meson production  $T_p\sim 0.9$ GeV one has  $|b_1|/|a_1|\sim 0.1$,
 therefore it allows one to put $R_0=1/3$ \cite {kuwilkf},\cite{kuwilkfpl}.
  Unfortunately, the experimental data on the spin
structure of the $pp\to d\pi^+$ and $\pi^+n\to \omega (\phi )p$ amplitudes
 at   energies $T_p\geq 1400 MeV$ are not available.
Thus,  the exact absolute magnitude of the spin factors and the cross sections
is rather questionable. We have found numerically from Eqs.(\ref{7}-
\ref{7a}) that the values $R_0$ and $R_1$ vary in the range from 1/9
to 4/9 when the complex amplitudes $a_i$ and $b_i$ vary
arbitrary.  An remarkable peculiarity of the condition $|a_1|\gg |b_1|$ is that
in this case the spin factor for vector mesons $R_1$  does not depend on the
 behaviour of amplitudes $a_2$ and $b_2$
and in accordance with
Eq.(\ref{7a}) equals to $R_1=1/3$. This value is very close to the maximal one
$R_S^{max}=
4/9$. As it will be shown below the assumption $|a_1|\gg |b_1|$, which provides
the condition $R_0=R_1={1\over 3}= const$, is compatible with main features
of the observed cross sections for $\eta, \omega$ and $\eta '$ meson
production.  The numerical calculations
 are present below at $R_0=R_1={1\over 3}$.

  The numerical calculations are performed using the RSC wave function of the
deuteron  \cite{alberi}. The parametrization  \cite {kuzuyu} of the overlap
integral between the three-body wave function of the $^3He$ nucleus and
the deuteron is used for the wave function  of
 $^3He$, $\Psi_{\tau }$, in the channel $d+p$. The value $S_{pd}^{\tau }=1.5$
is taken for the deuteron spectroscopic factor in  $^3He$ \cite {schiav}.
The numerical results are obtained in the $S$ - wave
 approximation for the spin averaged cross sections and taking into
account the D-component of deuteron for spin correlations.
The formfactor (\ref {2}) can be expressed through the S- and D-
components of the nuclei wave function $\varphi_l$ by the following integrals
\begin{equation}
{\cal F}_{Lll'}(P_0,E_0)={1\over 4\pi}\int _0^\infty j_L(P_0r)\exp{(iE_0r)}
\varphi^{\tau}_l(r)\varphi ^d_{l'}(r)r\ dr;
\label{8}
\end{equation}
 the normalization integral $\int _0^\infty [\varphi_0^2(r)+
\varphi_2^2(r)]r^2dr$ equals to 1 for the deuteron and $S_{pd}^\tau $
for the $^3He$.
 In the S-vawe
approximation we have ${\cal F}(P_0,E_0)={\cal F}_{0000}$.
The parametrization  \cite{ritchie} is used here for the differential cross
section of  the $pp\to d\pi^+$ reaction.
The experimental data on the total cross section of the reactions $\pi^+n\to
p\eta
(\eta',\omega,\phi) $ are taken from Ref. \cite {cern83} and the isotropic
 behaviour of the differential cross  section is assumed here.
In Fig.2 are shown the results of calculations of the  modulus squared of the
form factor  $|{\cal F}_{000}(P_0,E_0)|^2$ for the production of  $\eta,
\ \eta', \omega, \ \phi$ mesom at the angle  $180^0$ as a function of kinetic
 energy $T_p$ of the incident proton  in the laboratory
system. One can see from this figure that the value of $|{\cal F}_{000}
(P_0,E_0)|^2$  decreases exponentially with increasing
 $T_p$, and the slope in the logarithmic scale is the same for
all mesons in question. It is important to remark that at the definite energy
 $T_p$ the value of the form factor $|{\cal F}_{000}(P_0,E_0)|^2$ is
 practically  the same for all mesons  whose production thresholds in
 the reaction  $pD\to ^3HeX$  is below  $T_p$. Therefore difference in the
 production probability of different mesons in the two-step model is mainly
 due to  difference of the  $\pi^+n\to Xp$ amplitudes.

The results of calculations of the differential cross sections are presented
in Fig.3 in comparison with the experimental data  at $\theta_{c.m.}
=180^0$ from Refs.\cite{kusaturn},
 \cite {kuwurz95} and
at $\theta_{c.m.}=60^0$, $T_p=3$ GeV from Ref. \cite {kubrody}.
We do not discuss here the region near the $\eta $ threshold and the related
 problem of $\eta -^3He $ final state interaction which was investigated
in detail previously \cite {kuklu},\cite{kuwilkf}.
 It  follows from  Fig.3,{\it a} that form of energy dependence
 of the calculated cross section for the $pD\to ^3He \eta$ reaction
  at the energies sufficiently higher than the threshold
 $T_p\ge 1.3$ GeV ($p^*=0.4 - 1.0 $ GeV/c) is in qualitative agreement with
 the experimental data
 at  $\theta_{c.m.} =180^0$ and in a less degree at $\theta _{c.m.}=60^o$.
 To obtain the absolute value of the cross section we need the normalization
 factor $N=3$ which is close to $N=2.4$ found in Ref. \cite {kuwilkfpl} at
 the  threshold.
According to our calculations (Fig.3,{\it b}), the cross
section of  the $\eta '$ meson production near the threshold $(p^*=22$ MeV)
 and at $T_p=3$ GeV agrees with the experimental data in absolute value at
the same  factor $N=3$  as for the $\eta $ meson.
As one can see from Fig.4, the shape of the  modulus squared $|f|^2$ of the
 $pD\to ^3He\omega $ reaction amplitude as a function of momentum  $p^*$
 agrees properly with the  form observed in
experimental data \cite{kuwurz95} in the range of
 $p^*=0 - 500$ MeV/c. It should be noted that the ratio
 $R(\phi/\omega )=|f(pD\to ^3He \phi)|^2/|f(pD\to ^3He \omega)|^2$ near the
corresponding thresholds predicted by the model  $R^{th}=0.052$ is in
 agreement with the experimental value $R^{exp}=0.07\pm 0.02$.

  However, the
 absolute value of the cross section for vector mesons is essentially smaller
 than the experimental value. At the threshold ($p^*\sim  20 MeV/c$)
 the normalization factor N for $\omega $- is 5.9 and for $\phi $-meson is 6.6.
   To describe the absolute magnitude of the
cross section  in the range of $100 MeV/c\leq p^*\leq 400 MeV/c$ one needs
 the normalization factor $N=9.6$.
 On the other hand, at $T_p=3$ GeV for
  $\theta_{c.m.}=60^o$ \cite{kubrody}  the calculated cross section coincides
  with the experimental value in absolute magnitude.

 The above mentioned agreement with the experimental data in form of energy
 dependence of $\eta, \omega $  (and $\eta '$) and in
  the ratios $\eta'/\eta$,
  $\phi/\omega$  supports the assumption that the spin factors $R_0 $ and $R_1$
  are approximately constants in the corresponding energy regions.
   Therefore  the assumption $|a_1|\gg |b_1|$ seems to be enough reasonable.
   It allows us to give the definite prediction for spin-spin correlations
   in the reaction ${\vec p}{\vec D}\to ^3HeX$ with polarized deuteron and
   proton. Using Eqs.(\ref {6}, \ref{6a}) and taking into account the D-
   component of the deuteron wave function  we find under above condition
   $b_1=0$,
   that the cross section of vector meson production in case of
   polarized colliding particles can be obtained from  Eq.(\ref {5}) by the
    following    replacement
    $$  R_1|{\cal F}(P_0,E_0)|^2 \to    {1\over 3}\Biggl \{ |{\cal F}_{000}|^2
   (1- {\bf P}_p\cdot {\bf P}_D -\sqrt{2}Re({\cal F}_{000}{\cal F}_{202}^{\ *})
  \left [{\bf P}_p\cdot{\bf P}_D- {3\over {2{\bf P}_0^2}}
  [{\bf P}_0\times {\bf P}_p]\cdot
    [{\bf P}_0\times {\bf P}_D]\right ]$$
 \begin{equation}
   \label{11}
      +|{\cal F}_{202}|^2
    \left [1+{1\over 4}{\bf P}_p\cdot{\bf P}_D-{3\over {2{\bf P}_0^2}}
    [{\bf P}_0\times {\bf P}_p]\cdot[{\bf P}_0\times {\bf P}_D]
    +{9\over{ 4 {\bf P}_0^4}}
    ([{\bf P}_0\times {\bf P}_p]\cdot{\bf P}_0)([{\bf P}_0\times {\bf P}_D]
    \cdot{\bf P}_0)
    \right ] \Biggr \},
          \end{equation}
     where ${\bf P}_p$ and ${\bf P}_D$ are the polarization vectors of
   the proton and deuteron respectively and the momentum ${\bf P}_0$ is
    defined in Eq.(\ref{3}). It is assumed here that the tenzor
    polarization of deuteron is
  zero. Using  this result  we obtain for the spin-spin asymmetry the following
  expression  in case ${\bf P}_d\perp {\bf P}_0$ and ${\bf P}_p\perp {\bf P}_0$
   \begin{equation}
   \label{12}
   {\Sigma}_1= {d\sigma (\uparrow \uparrow)-d\sigma (\uparrow \downarrow)\over{
     d\sigma (\uparrow \uparrow)+d\sigma (\uparrow \downarrow)}}=
     -{{|{\cal F}_{000}|^2-|{\cal F}_{202}|^2-
     {1\over \sqrt{2}}Re({\cal F}_{000}{\cal F}_{202}^{\ *})}\over
     {|{\cal F}_{000}|^2+|{\cal F}_{202}|^2}},
  \end{equation}
  where $d\sigma (\uparrow \uparrow)$ and $d\sigma (\uparrow \downarrow)$
  are the cross sections for parallel
  and antiparallel orientation of the polarization vectors of the proton
  and deuteron.  We have found numerically from Eq. (\ref{12}) that near
   the threshold
  $\Sigma_1(\phi)=-0.95 $  and  $\Sigma_1$
   very fast goes to $-1$
  above the threshold.
  Very similar result is obtained for the $\omega $ meson: $\Sigma_1(\omega)
  =-0.92$. Neglecting the   D-component of
  the deuteron wave function we obtain one the same result for vector
  and pseudoscalar mesons:  $\Sigma_1=\Sigma_0=-1$.

 5. In conclusion,  the two step mechanism favoured  in the case of $\eta $
meson production near the threshold and at $\theta _{c.m.}\sim 90^o$ owing
to the kinematical velocity  matching  turns out to be very important also
beyond the matching conditions, namely both above the threshold
of $\eta $ meson production  and in the cases of  $\eta',\ \omega,$ and
$\phi $ mesons. Despite of its simplicity the two-step model describes fairly
well the shape of energy dependence of available experimental data
 of the cross
sections of $\eta$ and $\omega$ production as well as the ratios $\eta '/\eta $
and $\phi /\omega $ at the thresholds.
The absolute value of the cross sections   is not
described by this model. The most discrepancy was found for the vector mesons.
The normalization factor for $\omega $ meson $N= 9.6$ is considerably greater
than the value $N=2.4 $ established by F\"aldt and Wilkin \cite{kuwilkfpl} at
the thresholds of all mesons under discussion.
The reasons for the deficiency in absolute value of the predicted
cross section  may be
the contribution of nondeuteron states in the  subprocess $pp\to \pi NN$ at
the first step and the full spin structure of the elementary
amplitudes beyond the approximations (\ref{6},\ref{6a}) and (\ref{1}).
 For example, the contribution of the two-nucleon state with
the spin 0 will modify in a different way  the amplitudes of the pseudoscalar
and vector meson production in  the reaction $pD\to ^3HeX$.
 The experiments with polarized
particles \cite{kukhac} can give a new information
about the mechanism in question.

The authors are grateful to M.G.Sapozhnikov  for useful discussions.
This work was supported in part by grant $N^o$ 93-02-3745 of the Russian
Foundation For Fundamental Researches.

\newpage
\section*{Figure captions}
Fig.1.
Two-step mechanism of the reaction
$pD\to ^3HeX$.
\newline
Fig.2.
Calculated modulus squared  of form factor $|{\cal F}_{000}(P_0,E_0)|^2$
 (\ref {2},\ref {8})
as a function of kinetic energy of the proton in laboratory system
 $T_p$ for  $\eta ,\ \eta',\ \omega, \ \phi$ meson production
at $\theta_{c.m.}=180^0$.
\newline
Fig.3.
Differential cross sections of the $pD\to ^3He\eta (\omega ,\ \eta',\
\phi )$ reactions as a function of lab. kinetic energy of proton $T_p$.
The curves show the results of calculations at $R_S={1\over 3}$
for different angles $\theta _{c.m.} $ multiplied by the appropriate
normalization factor $N$.
\\
{\large {\it a}} - $pD\to ^3He\eta $:
   $180^\circ $ (full line, $N=3$), $90^\circ$ (dashed curve, $N=3$),
 circles  are experimental data:
 $\circ $ -
$\theta_{c.m.}=180^0$
Ref.{\protect \cite {berg88}} ; $\bullet $ -  $\theta_{c.m.}=60^0$
Ref.{\protect \cite {kubrody}};
\\
{\large {\it b}} - $pD\to ^3He\eta ' $ at $\theta_{c.m.}=180^0$ (full, N=3)
 and
$\theta_{c.m.}=60^0$ (dashed, N=3); the circles are experimental data for the
$\eta '$ production: $\circ $ -
$\theta_{c.m.}=180^0$
Ref. {\protect \cite {kusaturn}}; $\bullet $ -  $\theta_{c.m.}=60^0$
Ref.{\protect \cite {kubrody}}; the dotted line shows the results of
calculation
for the $pD\to ^3He\phi$ reaction at $\theta_{c.m.}=180^o$ normalized by factor
 $N=6.6$ to
 the experimental point ($\triangle $) from Ref.\cite{kusaturn};
\\
{\large {\it c}} - the same as {\large {\it b}}  but for the reaction  $pD\to
 ^3He\omega $ with normalization factor  $N=1$; circles ($\circ $) are
 the experimental data from Ref. {\protect \cite {kuwurz95}}.
\newline
Fig.4. The  modulus squared of the amplitude  of the $pD\to ^3He\omega $
        reaction defined by Eq.(\ref{4}) as a function of the c.m.s.
         momentum of the $\omega $ meson, $p^*$.
        The curve is the result of calculation at $R_1={1\over 3}$ multiplied
         by factor $N= 9.6$, the circles ($\circ $) are experimental data
         \cite{kuwurz95}.

\eject
\begin{figure}[t]
\label{fig1}
\centering
\mbox{\epsfig{figure=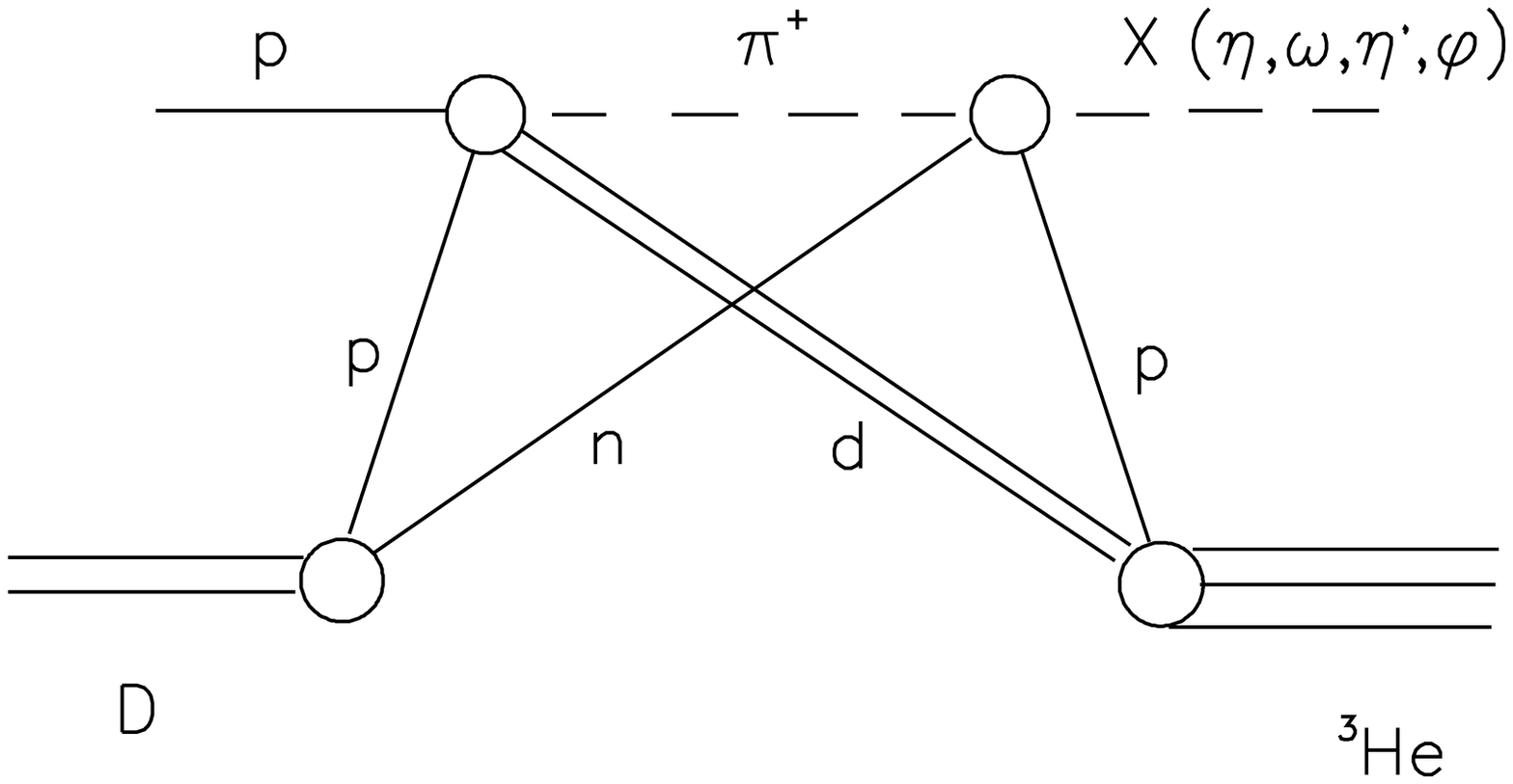,height=0.65\textheight, clip=}}
\caption{
 }

\end{figure}

\eject
\begin{figure}[h]
\label{fig2}
\centering
\mbox{\epsfig{figure=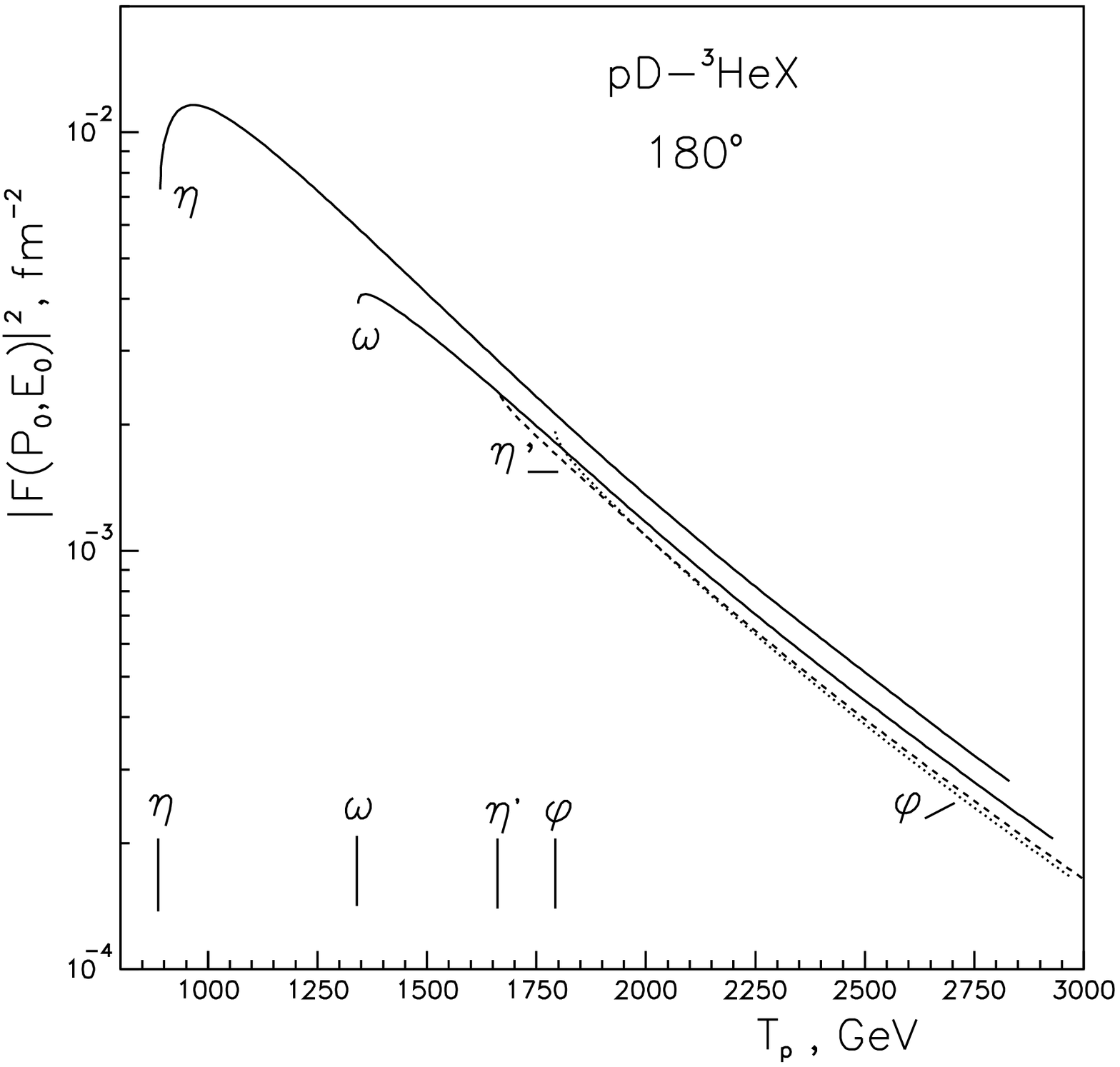,height=0.65\textheight, clip=}}
\caption {
}

\end{figure}

\eject
\begin{figure}[h]
\label{fig3}
\centering
\mbox{\epsfig{figure=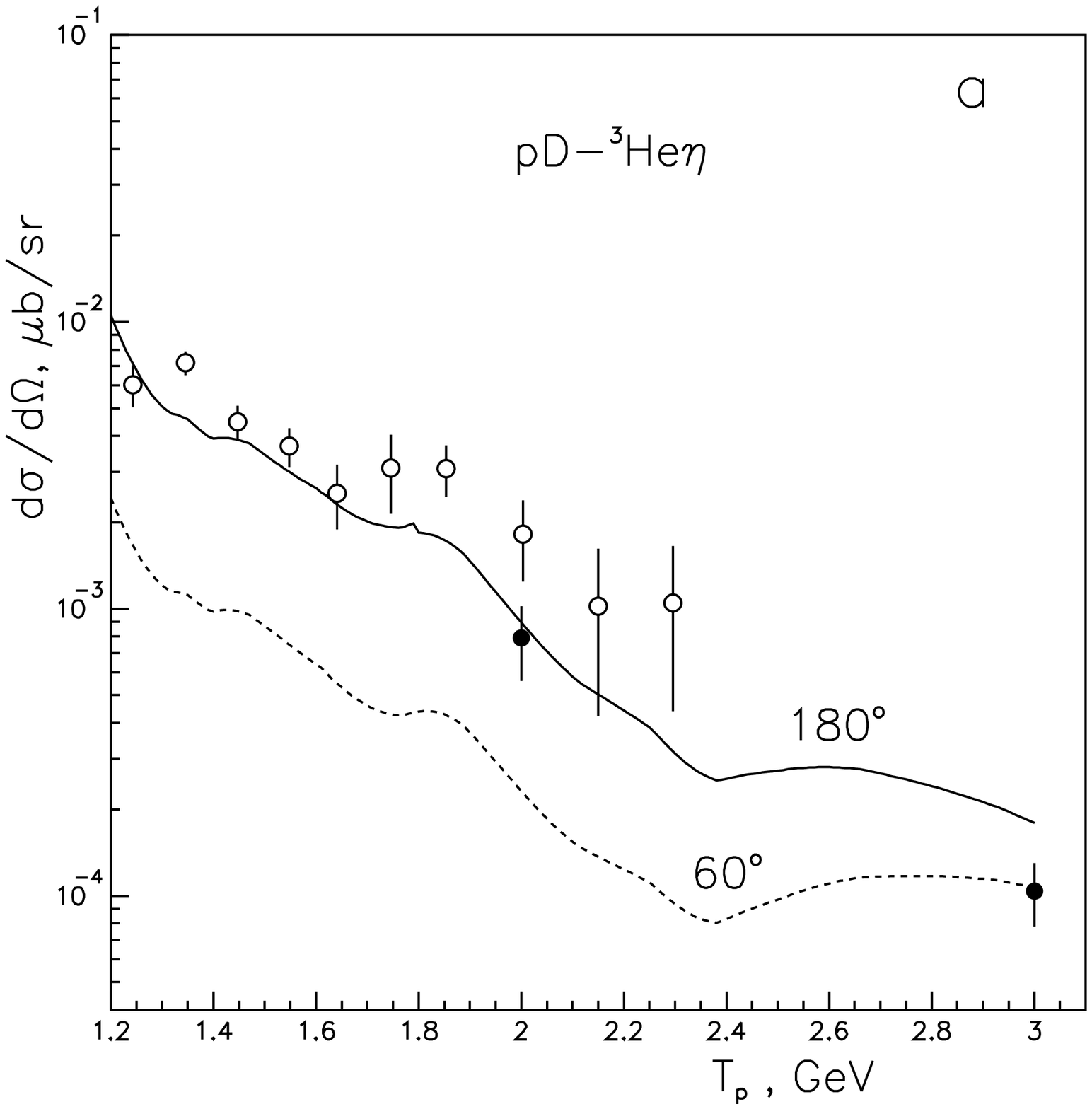,height=0.65\textheight, clip=}}
\centerline{FIG.3,a}
\end{figure}
\eject
\eject
\begin{figure}[h]
\centering
\mbox{\epsfig{figure=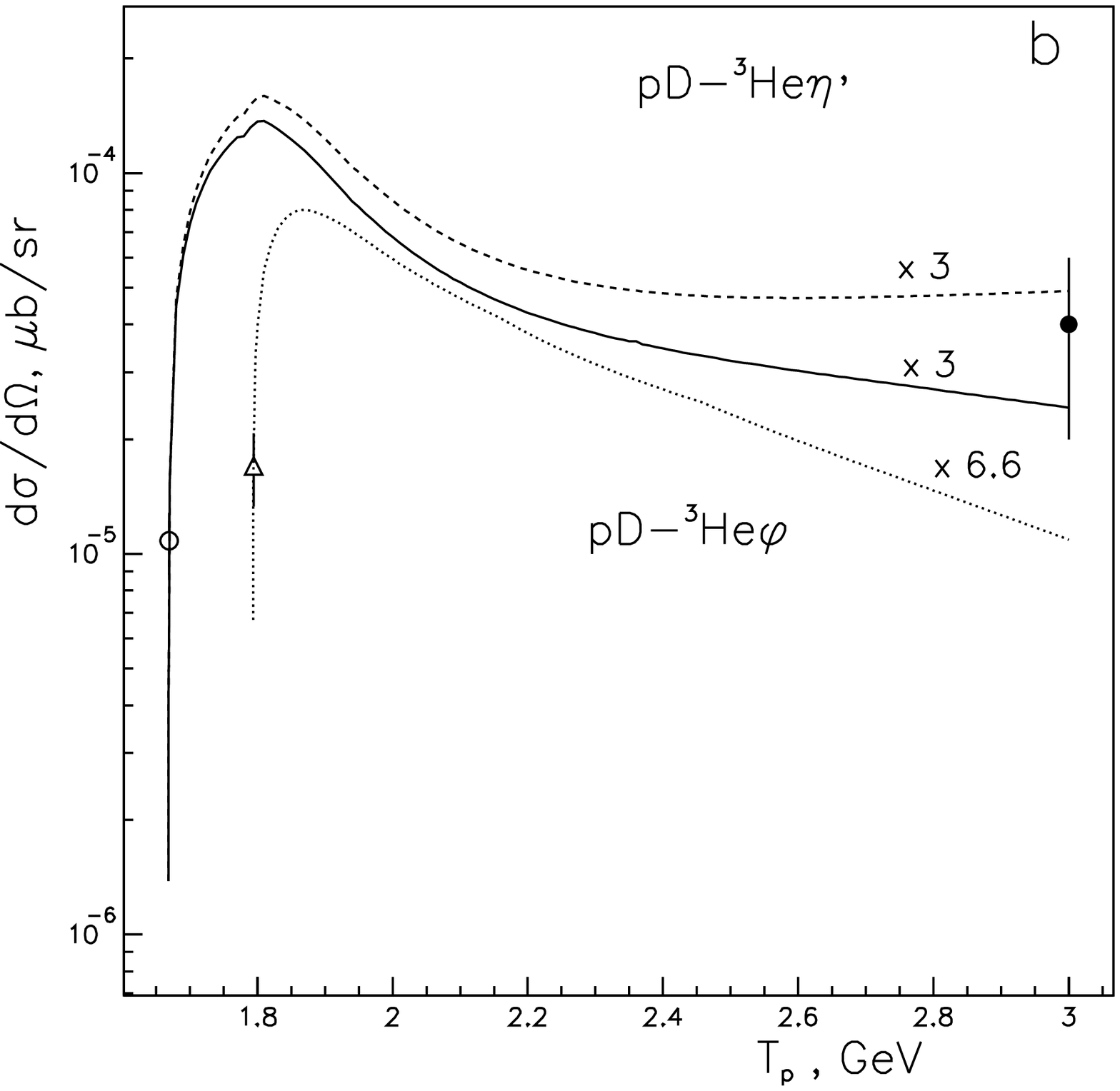,height=0.65\textheight, clip=}}
\centerline{FIG.3,b}
\end{figure}\eject
\begin{figure}[h]
\centering
\mbox{\epsfig{figure=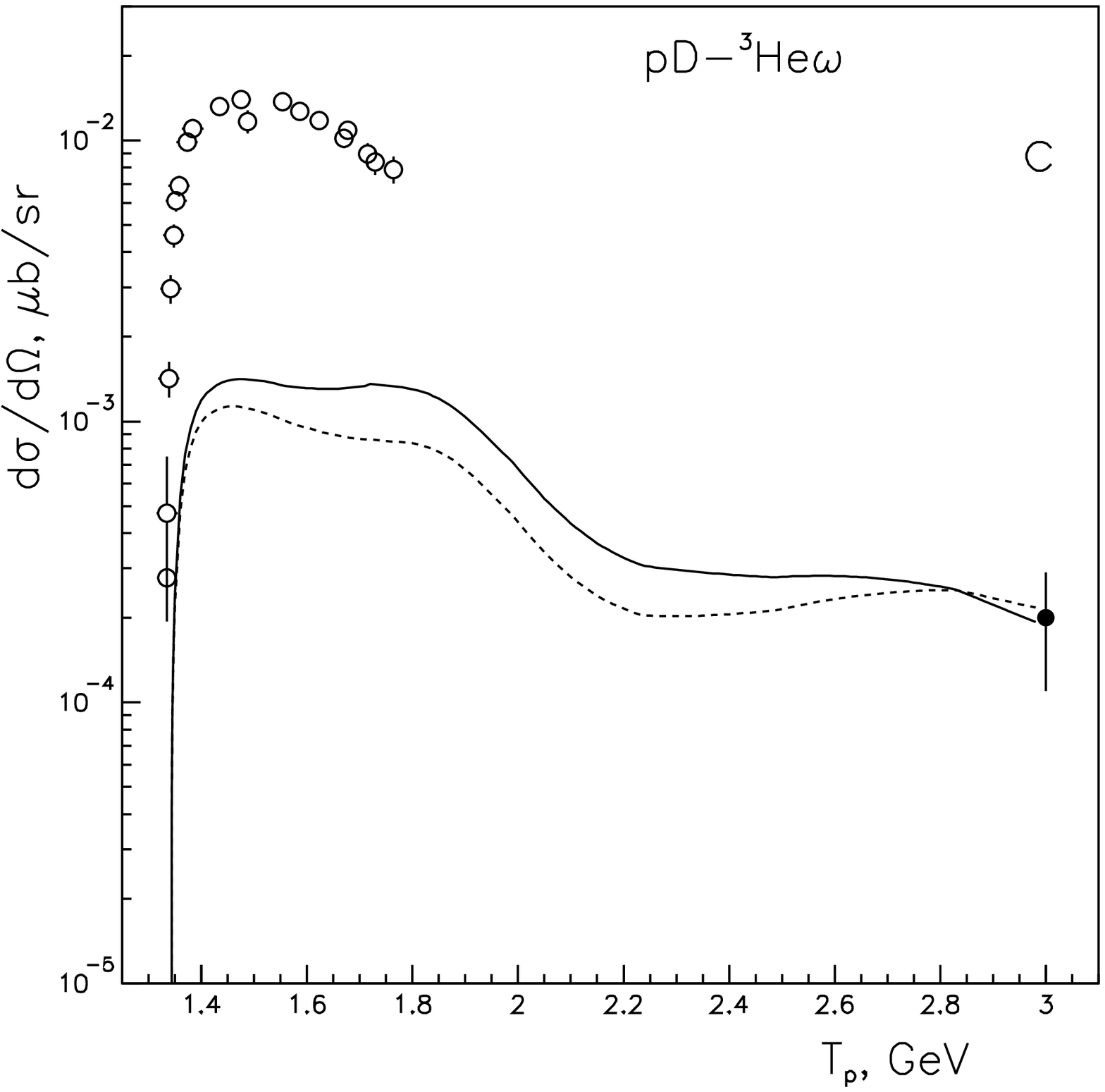,height=0.65\textheight, clip=}}
\centerline{FIG.3,c}
\end{figure}
\eject
\begin{figure}[h]
\centering
\mbox{\epsfig{figure=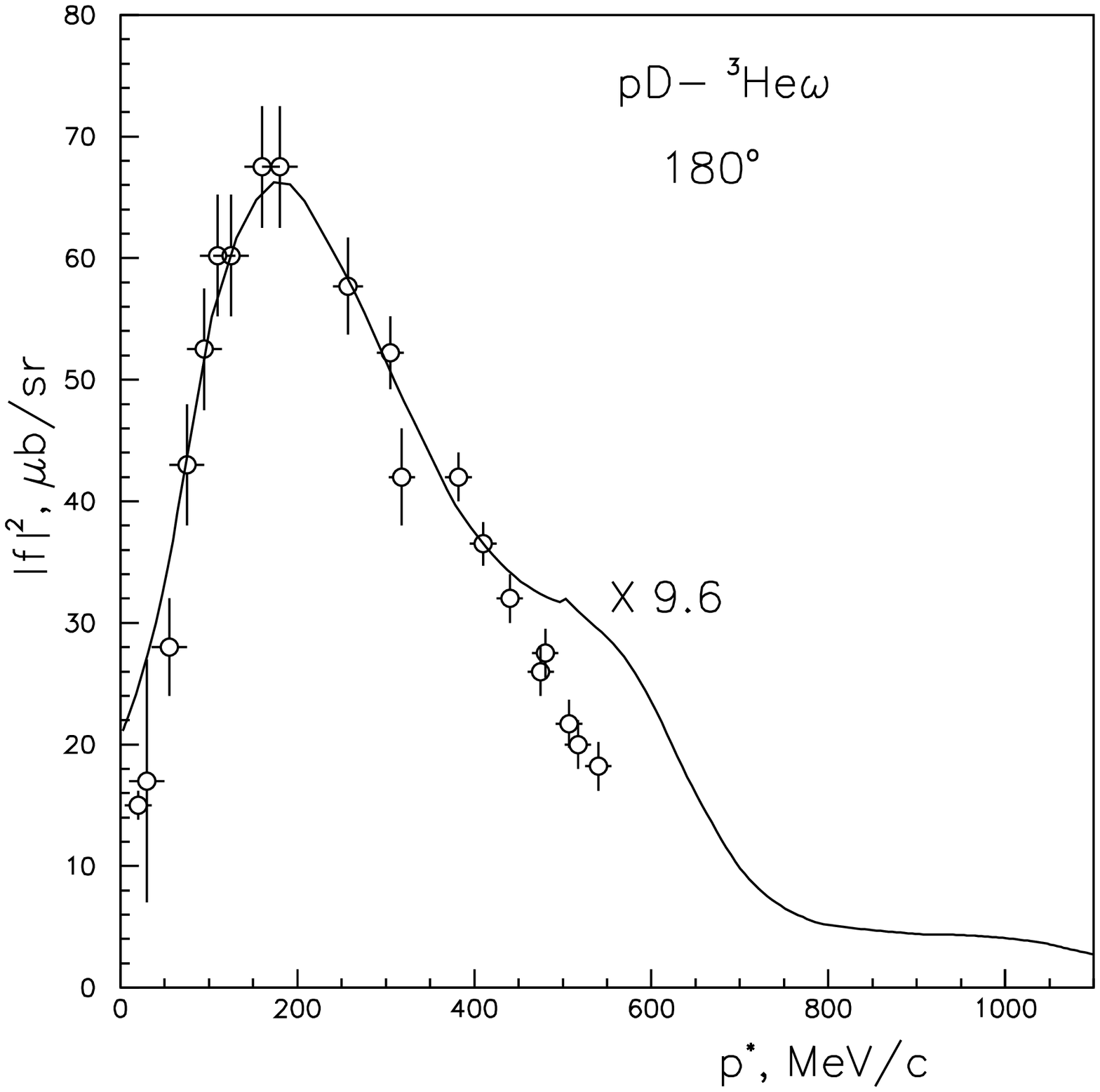,height=0.7\textheight, clip=}}
\centerline{Fig.4}
\end{figure}
\end{document}